\documentstyle[epsfig,aps,preprint]{revtex}
\begin{document}
\draft
\preprint{\vbox{\hbox{SOGANG-HEP 221/97}
                \hbox{hep-th/9706154} }}
\title{Graceful exit problem and stress-energy-momentum tensors
	revisited in the two-dimensional string cosmology}
\author{Won T. Kim\footnote{electronic address:wtkim@ccs.sogang.ac.kr}
  and Myung Seok Yoon\footnote{electronic address:younms@physics3.sogang.ac.kr}}
\address
{Department of Physics and Basic Science Research Institute, \\ 
Sogang University, Seoul 121-742, Korea }
\date{August 1997; revised}
\maketitle
\bigskip
\begin{abstract}
 We study the graceful exit problem and the role of the
stress-energy-momentum tensors in the two-dimensional string cosmology.
The one-loop quantum correction of conformal fields is incorporated
in the arbitrary large $N$ limit to ensure exact quantum
solvability. The only solution which gives the bounded curvature with
the asymptotic flatness is restricted to the first branch under some
conditions. However, even in this case, the accelerating expansion is
forever. We show that the only nonvanishing quantum stress-momentum
tensor is the pressure part($T_{xx}$) which is of relevance to the
dynamical evolution of the universe in the comoving coordinate. The
quantum energy part is zero since the negative contribution of the
induced conformal matters always cancels the positive quantity of the
induced dilaton part in terms of the constraint equation.
\end{abstract}
\bigskip

\newpage
 
 Recently, there has been much interests in the graceful exit
problem~\cite{bv} of the string inflationary cosmology~\cite{gv} with
the scale factor duality~\cite{vst}. The essential problem is due to
the curvature singularity arising from the classical cosmological
solutions of low-energy string theory. On the other hand, 
in the four-dimensional Einstein gravity
with the loop effect which is regarded as Einstein frame of string 
theory~\cite{gv2},
some cosmological singularity problems have been studied in terms of
the quantum back reaction of space time~\cite{art,em}. Further, the
quantum tunneling with the finite probability between the classically
distinct phases corresponding to the pre(accelerated) and
post(decelerating) big bang can be possible in the low energy string
background~\cite{gmv}. This fact may solve the graceful exit problem
of the string cosmology.

 The exactly soluble Callan-Gidding-Harvey-Strominger(CGHS)
model~\cite{cghs,rst}, which has been successful to deal with the
two-dimensional black holes, has been recently
investigated by Rey~\cite{rey} to show whether or not the
branch-changing phase transition appears in the two-dimensional string
cosmology.  This model was also extended to the generalized
two-dimensional dilaton gravity model 
by Gasperini and Veneziano~\cite{gv2d}.
The quantum back-reacted solution of the first branch
(second branch) with the bounded curvature scalar has been defined within
the whole comoving time from the past infinity to the future infinity, so that
the second branch (first branch) effectively disappear instead of connecting
two distinct branches. 

On the other hand, this model has the negative anomaly coefficient
which corresponds to the negative Hawking radiation and the number of
conformal matter fields are restricted to less than 24. Thus it is
natural to study the CGHS model in the manner to take the arbitrary
large positive number of matter fields to take good approximation within
the one-loop vacuum polarization of the conformal matter fields~\cite{bos}.  
Very recently, Bose and Kar~\cite{bk} suggest the way how to overcome 
the limit of $N$ by 
adding a local covariant counter-term and the exact scale factor
is obtained.      
    
 In this paper, we reconsider the graceful exit problem in the
two-dimensional string cosmology with the slightly generalized counter-terms 
containing the local term of Ref.~\cite{bk} and mainly study the role
of the stress-momentum tensors induced by the quantum corrections. The
only solution which gives the bounded curvature with the asymptotic
flatness is restricted to the first branch. However, even in this
case, the decelerating phase does not appear although the solution has
asymptotic flatness. We show that the only nonvanishing quantum
stress-energy-momentum tensor is the pressure part($T_{xx}$) which is
of relevance to the dynamical evolution of the universe in the
comoving coordinate. The total quantum-mechanical energy is zero since
the negative contribution of the induced conformal matters always
cancels the positive quantity of the induced dilaton part in terms of
constraint equation.

 Let us now consider the two-dimensional low-energy string theory given by
\begin{equation}\label{dg:act}
S_{{\rm{DG}}} = \frac{1}{2 \pi} \int\/d^{2}x\sqrt{-g}\, e^{-2\phi}
	\left[ R + 4(\nabla\phi)^2  + 4\lambda^2 \right],
\end{equation}
where $\phi$ is a dilaton field, and the cosmological constant
$\lambda^2$ sets to zero in that we are now considering dimensionally
reduced low-energy string theory from the critical dimensions.
The action for the classical and quantum
matter are written in the form of
\begin{eqnarray}
S_{{\rm{Cl}}} &=& - \frac{1}{2 \pi} \int\/d^{2}x\sqrt{-g}\,
	\frac12 \sum_{i=1}^N (\nabla f_i)^2, \label{cl:act}\\
S_{{\rm{Qt}}} &=& \frac{\kappa}{2\pi}\int d^2x\/ \sqrt{-g}\left[
	-\frac14 R\frac{1}{\Box}R + q\left(\nabla\phi\right)^2
	- \frac{\gamma}{2}\phi R\right],\label{qt:act-local}
\end{eqnarray}
where $\kappa = \frac{N-24}{12}$. 
In Eq.~(\ref{qt:act-local}), the
first term is induced matter part and the second and third represents
induced dilaton part. And $q$ and $\gamma$ simply denoted as $(q,
\gamma)$ are constants which will be chosen in later for exact solvability. 
For $(2, 6)$, the model was already treated by Bose and Kar in
Ref.~\cite{bk} and for $(0, 1)$ it is just the RST model which
corresponds to the Rey's model in the cosmology. For $(1,2)$, it is a
black hole model in Ref.~\cite{bpp}. For an arbitrary large $N$ ($N>24$),
we assume that the
anomaly coefficient $\kappa$ is finite~\cite{bk}. The nonlocal form of
the action (\ref{qt:act-local}) is written as by introducing an
auxiliary field $\psi$ for later convenience,
\begin{equation}
S_{{\rm{Qt}}} = \frac{\kappa}{2\pi}\int d^2x\/ \sqrt{-g}\left[
	\frac14 R\psi-\frac{1}{16}(\nabla\psi)^2
	+ q\left(\nabla\phi\right)^2
	- \frac{\gamma}{2}\phi R\right].\label{qt:nonlocal}
\end{equation}
The effective total action is 
\begin{equation}\label{eq:eff.act}
S_{\rm T} = S_{\rm DG} + S_{\rm M},
\end{equation}
where the matter part is composed of two pieces of $S_{\rm M}=S_{\rm
Cl} + S_{\rm Qt}$. The equations of motion and the constraint
equations for this action (\ref{eq:eff.act}) are  
\begin{equation}
\label{eom}
G_{\mu\nu} = T_{\mu\nu}^{\rm M}
\end{equation} 
where
\begin{eqnarray}
& &G_{\mu\nu} = \frac{2\pi}{\sqrt{-g}}\frac{\delta S_{\rm DG}}
	{\delta g^{\mu\nu}} \nonumber \\
& & \qquad \/ =e^{-2\phi}\bigg[ 2\nabla_\mu\nabla_\nu\phi
	+ 2g_{\mu\nu}\left((\nabla\phi)^2-\Box\phi
	\right)\bigg],\label{cov:G} \\
& &T_{\mu\nu}^{\rm M} \ = -\frac{2\pi}{\sqrt{-g}}
	\frac{\delta S_{\rm M}}{\delta g^{\mu\nu}} \nonumber \\
& & \qquad \, =\frac12 \sum_{i=1}^N \left[\nabla_\mu f_i \nabla_\nu
	f_i -\frac{1}{2}(\nabla f_i)^2 \right] +
	\frac{\kappa}{4}\left[\nabla_\mu\nabla_\nu\psi +
	\frac14\nabla_\mu\psi\nabla_\nu\psi -
	g_{\mu\nu}\left(\Box\psi+\frac18(\nabla\psi)^2 \right)\right]
	\nonumber \\
& & \qquad \ \ \ -\frac{\gamma\kappa}{2}\bigg[
	\nabla_\mu\nabla_\nu\phi - g_{\mu\nu}\Box\phi\bigg]
	-q\kappa\left[\nabla_\mu\phi\nabla_\nu\phi
	- \frac12 g_{\mu\nu}(\nabla\phi)^2\right],
	\label{cov:T} 
\end{eqnarray}
and the other equations of motion for $\phi$, $f_i$, and $\psi$
are given by
\begin{eqnarray}
& & e^{-2\phi}\left[R+4\Box\phi-4(\nabla\phi)^2 \right]
	 = -\frac{\gamma\kappa}{4}R-q\kappa\Box\phi,\label{cov:phi}\\
& &\Box f_i =0, \label{cov:matt} \\
& &\Box \psi = -2R. \label{cov:psi}
\end{eqnarray}
In the conformal gauge, $g_{\pm\mp}=-\frac12 e^{2\rho}$,
$g_{\pm\pm}=0$, the total action and the constraints are given by
\begin{eqnarray}
S_{\rm T} &=& \frac{1}{\pi}\int\/d^2 x \bigg\{ 
	e^{-2\phi}\left[ 2\partial_+\partial_-\rho
	- 4\partial_+\phi\partial_-\phi
	\right] 
	- \kappa\Big[ \partial_+\rho\partial_-\rho 
	+\gamma\phi\partial_+\partial_-\rho \nonumber \\
  & & + q \partial_+\phi\partial_-\phi \Big]
	+ \frac12 \sum_i \partial_+f_i\partial_-f_i
	\bigg\}\label{conf:act}
\end{eqnarray}
and
\begin{eqnarray}
& &e^{-2\phi}\left[ 4\partial_\pm\rho\partial_\pm\phi
	- 2\partial_\pm^2\phi \right]
	+ \frac12\sum_{i=1}^N\left(\partial_\pm f\right)^2
	+ \kappa\left[ \partial_\pm^2\rho
	- \left(\partial_\pm\rho\right)^2\right]\nonumber \\
& &\qquad\qquad\qquad\qquad\qquad - \frac{\gamma\kappa}{2}\left(
	\partial_\pm^2\phi - 2\partial_\pm\rho\partial_\pm\phi\right)
	- q\kappa\left(\partial_\pm\phi\right)^2
	- \kappa t_\pm = 0, \label{conf:constr}
\end{eqnarray}
where $t_{\pm}$ reflects the nonlocality of the induced gravity of
the conformal anomaly~\cite{wl}and we set $q=\gamma-1$ to make the
equations exactly solvable. Without the classical matter, $f_i = 0$,
defining new fields as follows~\cite{wl,bc,dea},
\begin{eqnarray}
\Omega &=& -\frac{\kappa}{2}(\gamma - 2)\phi
	+ e^{-2\phi}, \label{qt:Omega} \\
\chi &=& \kappa\rho - \frac{\gamma\kappa}{2}\phi
	+ e^{-2\phi},\label{qt:chi}
\end{eqnarray}
the gauge fixed action is obtained in the simple form of
\begin{equation}
S_{\rm T} = \frac{1}{\pi} \int\/d^2 x \left[
	\frac{1}{\kappa}\partial_+\Omega\partial_-\Omega
	-\frac{1}{\kappa}\partial_+\chi\partial_-\chi 
	+ \frac12\sum_{i=1}^N\partial_+f_i\partial_-f_i \right]
\end{equation}
and the equations of motion and the constraints are given by
\begin{equation}
\partial_+\partial_-\Omega = \partial_+\partial_-\chi = 0, 
\end{equation}
\begin{eqnarray}
G_{\pm\pm} -T_{\pm\pm}^{\rm M} &=& -\frac{1}{\kappa}\left(
	\partial_\pm\Omega\right)^2 + \frac{1}{\kappa}
	\left(\partial_\pm\chi\right)^2 - \partial_\pm^2\chi
	- \frac12\sum_{i=1}^N(\nabla f_i)^2
	+ \kappa t_\pm \nonumber \\
	&=& 0. \label{qt:constr.}
\end{eqnarray}

In the homogeneous condition of fields, 
we obtain equations of motion in the simple forms of
\begin{equation}\label{eq:mot-t}
\ddot{\chi}(t) = \ddot{\Omega}(t) = 0,
\end{equation}
and they yield solutions,
\begin{eqnarray}
\chi &=& \kappa\rho - \frac{\gamma\kappa}{2}\phi + e^{-2\phi}
	= \chi_0 t + A, \label{sol:chi}\\
\Omega &=& \frac{\kappa}{2}\left(2-\gamma\right)\phi
	+ e^{-2\phi} = \Omega_0 t + B, \label{sol:om}
\end{eqnarray}
where $\chi_0,~~\Omega_0,~~A,$ and $B$ are constants.
The constraint (\ref{qt:constr.}) becomes, by using the solutions
(\ref{sol:chi}) and (\ref{sol:om}),
\begin{equation}
\kappa t_\pm-\frac{1}{4\kappa}\left(\Omega_0
	- \chi_0\right)\left(\Omega_0 + \chi_0\right) = 0.
\end{equation}
Choosing the quantum matter state as vacuum, $t_\pm =0$,
the first branch which corresponds to the case with
$\Omega_0 = \chi_0 \equiv - M (<0)$ and $A=B=0$ is obtained~\cite{rey}. 

At this juncture, let us study the boundedness of the scalar curvature
which is given by
\begin{eqnarray}
R &=& 2e^{-2\rho}\ddot{\rho}(t) \nonumber \\
  &=& M^2 e^{-2\rho}e^{-2\phi}\left[e^{-2\phi}
	+ \frac{\kappa}{4}(\gamma-2)\right]^{-3},
\end{eqnarray}
in terms of the explicit solutions (\ref{sol:chi}) and
(\ref{sol:om}). It is natural to confine the constant as $\gamma >2$
to avoid the singularity of the curvature. If $\gamma<2$, $\kappa$ should
be negative~\cite{wy}.
The solution (\ref{sol:chi}) and (\ref{sol:om}) belongs to the two branches
depending on the parameters.
Note that as far as we are concerned with the first branch solution, 
the accelerating expansion is only possible 
since without any assumption, the curvature is positive definite
$\ddot{a} >0$ in the whole range of the conformal time (or comoving
time) in our case. In fact,  even for the negative $\kappa$ which
corresponds to Rey's model, there is no decelerating phase~\cite{wy}.

On the other hand, we are concerned with the stress-energy-momentum tensors
obtained from (\ref{cov:T}),
\begin{eqnarray}
T_{\pm\pm}^{\rm M} &=& \kappa\left[\partial_\pm^2\rho
	- (\partial_\pm\rho)^2 - t_\pm(x^\pm)\right]
	- \frac{\gamma\kappa}{2}\left[\partial_\pm^2\phi
	- 2\partial_\pm\rho\partial_\pm\phi\right]
	- \kappa(\gamma-1)(\partial_\pm\phi)^2, \label{T:++} \\
T_{+-}^{\rm M} &=& -\kappa\partial_+\partial_-\rho +\frac{\gamma\kappa}{2}
         \partial_+\partial_- \phi.
	\label{T:+-}
\end{eqnarray}
By using the relation, $\rho(t) = \ln a(\tau)$ with $dt=d\tau/a(\tau)$,
one can perform the coordinate transformation to the comoving time,
and then Eqs. (\ref{T:++}) and (\ref{T:+-}) are 
\begin{eqnarray}
T_{\tau\tau}^{\rm M}(\tau) &=& 0, \label{T:+tt}\\
T_{\tau x}^{\rm M} (\tau) &=& 0, \label{T:+tx}\\
T_{xx}^{\rm M}(\tau) &=& 
       - \frac{2M^2\left[M\tau +\sqrt{(M\tau)^2
	+4\kappa(\gamma -2)}~\right]}{\left[(M\tau)^2 +4\kappa(\gamma
	-2)\right]^{\frac{3}{2}}}. \label{T:+xx}  
\end{eqnarray}
Note that the energy $T_{\tau\tau}^{\rm M}$ and the momentum $T_{\tau
x}^{\rm M}$ vanish and $T_{xx}^{\rm M}$ has a negative definite. 
As a result, the induced energy-momentum tensors are zero at anytime
while the stress part which corresponds to the pressure of the prefect
fluid has a time-dependent negative value. This fact seems to be
unusual and we clarify in detail in the followings.

 From now on, we choose comoving coordinates on the purpose
of directly computing the stress-energy-momentum tensors instead
of transforming from the results in the conformal gauge to the
comoving coordinates. The energy-momentum tensors are not true tensor
generically in the general coordinate transformation, for example, for
the conformal transformation in the black hole geometry~\cite{wl}. So
we now calculate the expectation value of the stress-energy-momentum
tensors from the beginning in the comoving coordinates given by
\begin{equation}\label{metric:t}
ds^2 = - d\tau^2 + a^2(\tau) dx^2.
\end{equation}
Then the classical dilaton gravity part $G_{\mu\nu}$ is written as
\begin{eqnarray}
& &G_{\tau\tau}(\tau) = 2 e^{-2\phi}\left[\dot{\phi}^2
	- \frac{\dot{a}}{a}\dot{\phi}  \right],
	\label{G:tt} \\
& &G_{\tau x}(\tau) =0, \label{G:tx} \\
& &G_{xx}(\tau) = 2 a^2 e^{-2\phi}\left[\ddot{\phi} - \dot{\phi}^2
	 \right], \label{G:xx} 
\end{eqnarray}
and the stress-energy-momentum tensors are, respectively,
\begin{eqnarray}
& &T_{\tau\tau}^{\rm M}(\tau) = \frac14 \sum_{i=1}^N \dot{f}_i^2
	- \frac{\kappa}{2}\left(\frac{\dot{a}}{a}\right)^2
	+ \frac{\gamma\kappa}{2}\frac{\dot{a}}{a}\dot{\phi}
	- \frac12 q\kappa\dot{\phi}^2 - \kappa t_{\tau\tau},
	\label{T:tt} \\
& &T_{\tau x}^{\rm M}(\tau) = 0, \label{T:tx} \\
& &T_{xx}^{\rm M}(\tau) = a^2 \left\{ \frac14\sum_{i=1}^N\dot{f}_i^2
	+ \kappa\left[\frac{\ddot{a}}{a}
	- \frac12\left(\frac{\dot{a}}{a}\right)^2\right]
	- \frac{\gamma\kappa}{2}\ddot{\phi}
	-\frac12q\kappa\dot{\phi}^2\right\} - \kappa t_{xx},
	\label{T:xx} 
\end{eqnarray}
where the overdots denote the differentiation with respect
to the comoving time $\tau$. The Eqs. (\ref{cov:phi}), (\ref{cov:matt}),
and (\ref{cov:psi}) are written in the form of
\begin{eqnarray}
& &e^{-2\phi}\left(\frac{2\ddot{a}}{a} + 4\dot{\phi}^2 - 4\ddot{\phi}
	- 4\frac{\dot{a}}{a}\dot{\phi} \right)
	= -\frac{\gamma\kappa}{2}\frac{\ddot{a}}{a}
	+ q\kappa\left(\ddot{\phi} +
	\frac{\dot{a}}{a}\dot{\phi}\right), \label{t:phi} \\
& &\ddot{f}_i + \frac{\dot{a}}{a}\dot{f}_i = 0, \label{t:f}\\
& &\ddot{\psi} + \frac{\dot{a}}{a} \dot{\psi}=4 \frac{\ddot{a}}{a}
	\label{com:psi}.
\end{eqnarray}
By eliminating the auxiliary field
$\dot{\psi}$, the integration ambiguities $t_{\tau\tau}(\tau)$ and 
$t_{xx}(\tau)$ were obtained as $a^2 \kappa t_{\tau\tau} = \kappa t_{xx} =
-\frac{\kappa}{32}C^2$, where C is an arbitrary constant. 
Thus $t_{\tau\tau}$ and $t_{xx}$ reflect the nonlocality of the
effective action. To find an exact solution, we set $f_i=0$,
$q=\gamma-1$, and $C=0$. The constraint equation $G_{\tau\tau} -
T_{\tau\tau}^{\rm M} = 0$ from Eqs. (\ref{G:tt}) and (\ref{T:tt}) is
neatly expressed as
\begin{eqnarray}
G_{\tau\tau} - T_{\tau\tau}^{\rm M} &=& -(\dot{\phi}
	- \frac{\dot{a}}{a})\left[-2\dot{\phi}e^{-2\phi}
	+ \kappa\left(\frac{1+\gamma}{2}\dot{\phi}
	+ \frac12\frac{\dot{a}}{a}\right)\right] \nonumber \\
	&=& 0, \label{t:constr}
\end{eqnarray}
and we choose $\dot{\phi} =  \frac{\dot{a}}{a}$, corresponding to the
first branch in Ref.~\cite{rey,bk}. Then note that
$T_{\tau\tau}^{\rm{M}}(\tau)$ is zero as far as we are concerned with
the first branch, which is easily shown by Eq.~(\ref{T:tt}). On the
other hand, the dynamical equation of motion from Eqs.~(\ref{G:xx})
and (\ref{T:xx}) is
\begin{equation}
\label{G=T}
G_{xx} - T_{xx}^{\rm M} =0,
\end{equation}
which yields
\begin{equation}\label{sol:a}
\frac{\kappa}{2}\left(\frac{\gamma}{2}-1\right)a(\tau) - \frac{1}{a(\tau)}
	= \alpha\tau + \beta,
\end{equation}
where $G_{xx} = 2\left[\frac{\ddot{a}}{a}
-2\left(\frac{\dot{a}}{a}\right)^2\right]$ and $T_{xx}^{\rm M} =
-\frac{\kappa}{2}(\gamma - 2)a\ddot{a}$ in terms of scalar
factor. Note that if $\alpha$ is fixed as $\frac12 M$ and the integration
constant $\beta$ is zero, then $T_{\tau\tau}^{\rm M}(\tau)$ and
$T_{xx}^{\rm M}(\tau)$ are exactly given by Eqs. (\ref{T:+tt}) and
(\ref{T:+xx}). From the above solution (\ref{sol:a}), the scale factor
and the curvature are given by
\begin{eqnarray}
a(\tau) &=& \frac{2}{\kappa(\gamma-2)}\left\{\alpha\tau +
	\sqrt{(\alpha\tau)^2 + \kappa(\gamma-2)}\right\},
	\label{scale:t}\\
R(\tau) &=& \frac{2\kappa(\gamma-2)\alpha^2}{\left[\alpha^2\tau^2 +
	\kappa(\gamma-2)\right]^{\frac32} \left[ \alpha\tau +
	\sqrt{(\alpha\tau)^2 + \kappa(\gamma-2)}\right]},
	\label{R:t}
\end{eqnarray}
where these are essentially same with those of Bose and Kar in Ref.~\cite{bk}.
In fact, the energy-momentum tensors $T_{\pm\pm}^{{\rm M}}$
(or $T_{\tau\tau}^{{\rm M}}$) are composed of the induced conformal
matter density and the dilaton part, however, they are exactly
canceled, which is shown by direct calculation of
Eq. (\ref{T:tt}). So, Eqs. (\ref{T:+tt}), (\ref{T:+tx}), and
(\ref{T:+xx}) are exactly reproduced from Eqs. (\ref{T:tt}),
(\ref{T:tx}),and (\ref{T:xx}) without considering an anomalous
transformation. Therefore, the dynamical evolution of the back-reacted
geometry in this cosmology is due to the quantum-mechanically induced
shear from Eq. (\ref{G=T}). Our energy density is parted as a matter
of convenience in two contributions from the nonlocal effective action
and some local ambiguity parts. In fact, the split of energy momentum
density is arbitrary so that the total energy momentum density is in
fact meaningful. And this null value can be changed by the addition of
the classical energy of the conformal fields or choosing the different
boundary condition $t_{\pm}$.

\begin{figure}
%\hglue1cm
\vspace{-3cm}
\begin{center}
\epsfig{figure=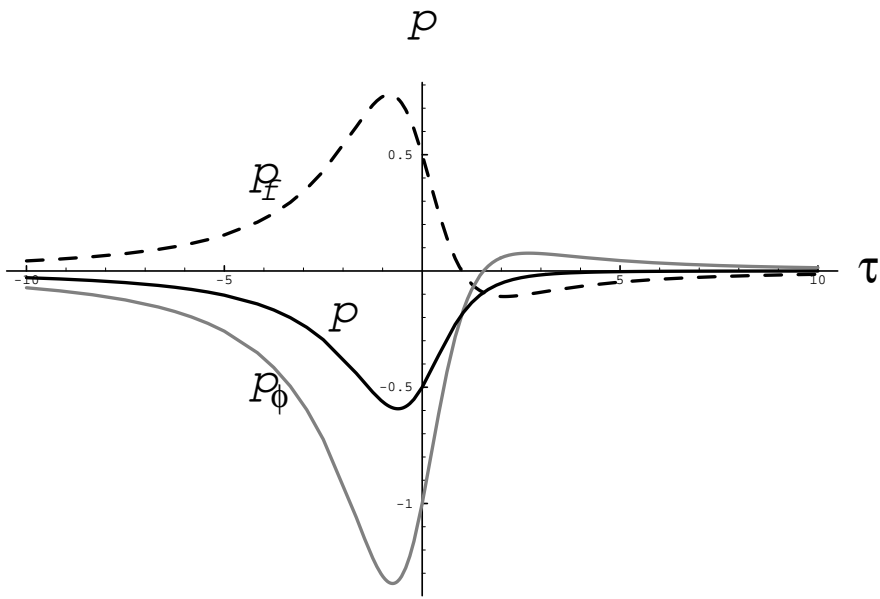,height=15cm}
\end{center}
\vspace{-2.5cm}
\caption{$p_f$ and $p_\phi$ have not definite sign, however the total pressure $p$ is negative definite.}
\label{fig:pressure}
\end{figure}

As a comment, it is easily shown that for $\tau\rightarrow -\infty$,
$a(\tau)\rightarrow -\frac{1}{\alpha\tau}$ and $R\rightarrow
\left(\frac{2}{-\tau}\right)^2$ (superinflation phase). In the limit
of far future, $\tau\rightarrow +\infty$, $a(\tau) \sim \alpha\tau$
and $R\sim\frac{1}{(\tau)^2}$ which corresponds to the Milne universe
(flat space-time). In the above analysis,
the universe keeps on accelerating expansion since $\dot{a}(\tau)>0$
and $\ddot{a}(\tau) > 0$ with the bounded curvature. The expected
decelerating phase~\cite{gmv} in the string cosmology does not appear
even in the Rey's model~\cite{rey}.

If we assume the induced quantum matter as a perfect fluid,
then the stress-energy-momentum tensors are 
\begin{equation}\label{T:p-rho}
T_{\mu\nu}^{\rm M} = p g_{\mu\nu} + (p+\rho)u_\mu u_\nu,
\end{equation}
where $u_\mu = (1,0)$, and $p$ and $\rho$ are pressure and energy
density, respectively. Then the only nonvanishing tensor is the shear
part and is written in the form of
\begin{eqnarray}
p &=& \frac{1}{a^2}T_{xx}^{\rm M}, \nonumber \\
  &=& -\frac{\alpha^2\kappa^2(\gamma-2)^2}{2\left[(\alpha\tau)^2
	+\kappa(\gamma-2)\right]^{\frac32}\left[\alpha\tau
	+\sqrt{(\alpha\tau)^2 +\kappa(\gamma-2)}\right]}.
	\label{p}
\end{eqnarray}
The matter and dilaton contribution to the pressure can be worked out
separately, and then they have not definite sign (see
Fig.\ref{fig:pressure}). The pressure $p_f$ from the induced matter
and $p_{\phi}$ from the induced dilaton part in Eq. (\ref{cov:T})
are explicitly given by 
\begin{eqnarray}
p_f &=& \kappa\alpha^2 \left[(\alpha\tau)^2
	+\kappa(\gamma-2)\right]^{-\frac32} \left[ -\alpha\tau
	+\frac12\sqrt{(\alpha\tau)^2 +\kappa(\gamma-2)}\right],
	\label{press:f} \\
p_\phi &=& \kappa\alpha^2 \left[(\alpha\tau)^2
	+\kappa(\gamma-2)\right]^{-\frac32} \left[
	\frac{\gamma}{2}\alpha\tau
	+\frac{1-\gamma}{2}\sqrt{(\alpha\tau)^2 +\kappa(\gamma-2)}
	\right]. \label{press:phi}
\end{eqnarray}
However, the total pressure is always
negative definite and depends on time. Hence one can state even though
there is no total energy density as a source, the pressure of the
radiation fields (shear) by the quantum effect appears to make the curvature
finite.

Now, it seems to be appropriate to discuss the positivity of $\kappa$.
It might be possible formally to take the negative limit $N \rightarrow
-\infty$ to obtain the good approximation if one restricts $N<24$ like
in Ref.~\cite{rey}. However, there may be some reasons why we have taken
the physical condition $N>24$ or $\kappa>0$. Essentially it arises
from the conformal field theory and consistency condition of locally
symmetric theory. First, from the conformal field theory point of
view, the central charge of matter field $c_{matter}=N$ should be
positive definite for the unitary theory. In other words, in a
nonunitary theory, the central charge is negative. So the number $N$
should be positive definite. As a second step, the gravitational field
turn on, the total central charge in our case is shifted by
$c_{total}=c_{matter} +c_{ghost} +c_{gravity}=N - 26+1+1-12\kappa =
N-24-12\kappa$. Up to this stage, $\kappa$ is still arbitrary. The
total central charge should be zero in order to preserve the
consistency of the theory, then $\kappa=\frac{N-24}{12}$. So the
large $\kappa$ or $N$ limit can be possible when $N>24$ as far as
we take $N>0$. On the other hand, the restriction of $N>24$ in our
theory from the regularity of curvature scalar also seems to be also
awkward. This can be improved by adding the Strominger's ghost decoupling
term~\cite{str} to the quantum effective action by using the
regularization ambiguity, then $\kappa=\frac{N}{12}$ is
obtained. Therefore, neither $N>24$ or $N<24$ may be necessary.

Two final remarks are in order. First, we did not consider the anomalous
transformation or Schwartzian derivative between the two coordinates
(conformal and comoving coordinates). As easily seen from the results
in the above two coordinates, there is no anomalous transformation of
stress-energy-momentum tensors in our cosmological model. So the
stress-energy-momentum tensors in the conformal gauge was transformed
to the comoving coordinate without any anomaly and result in the
vanishing energy-momentum tensors. This result is compatible with the
direct calculations in the comoving coordinate. So the integration
ambiguity $t_{\pm\pm}(t)=0$ was maintained in the form
$t_{\tau\tau}(\tau)=t_{xx}(\tau)=0$. In the black hole case, the
anomalous transformation of stress-energy-momentum tensors was assumed
to be canceled by the anomalous transformation of the integration
ambiguity, $t_{\pm\pm}(x^+,x^-)$.  Secondly, the covariant
conservation of $T_{\mu\nu}^{M}$ is spoiled by the local counter
term. In fact, since we are in the string-frame, the classical dilaton
gravity part is not purely geometrical on the contrary to the Einstein
gravity in the four dimensions. So there does not exist the covariant
conservation of the matter part although the whole tensor
$G_{\mu\nu}-T_{\mu\nu}^{\rm M}=0$ is covariant. To make this explicit,
by using the Eqs. (\ref{T:++}) and (\ref{T:+-}), one obtains
$\nabla_{\mu}T^{\mu\pm}_{\rm
M}=4\kappa(\gamma-2q)e^{-4\phi}\partial_{\mp}\phi
\partial_{\pm}\partial_{\mp} \phi$. Therefore, the
stress-energy-momentum tensor of induced quantum matter is not
covariantly conserved unless $\gamma=2q$. On the other hand, we
assumed $q=\gamma -1$ for exact solvability. So the covariant
conservation of matter part and exact solubility of the closed form
require $(1,2)$ called BPP model~\cite{bpp}, however, the curvature
scalar is not bounded.

 In summary, we have studied the superinflation of pre-big bang phase
is smoothly connected to flat universe 
in the large $N$ limit in the CGHS model. As far as we are concerned
with the vacuum theory, there is neither energy nor momentum while the
pressure (shear) of the radiation field governs the universe  in terms
of the quantum back reaction in this model.

{\bf Acknowledgments}

This work was supported by Ministry of Education, 1997, Project No. 
BSRI-97-2414, and Korea Science and 
Engineering Foundation through the Center for Theoretical Physics
in Seoul National University(1997).

%%%%%%%%%%%%%%%%%%%% References %%%%%%%%%%%%%%%%%%%%%%%%%

\end{document}